%% file: text.tex
\documentclass[manuscript]{aastex}
\usepackage{natbib}
\usepackage{color}

\shorttitle{Variability and TeV $\gamma$-ray duty cycle of Mrk 421 as determined by 3-Year of Milagro monitoring}
\shortauthors{Abdo et al.}

\usepackage{lineno}

\begin{document}

\title{The Study of TeV Variability and Duty Cycle of Mrk 421 from 3 Years of Observations with the Milagro Observatory}

\input{authorlist.tex}
\begin{abstract}
TeV flaring activity with time scales as short as tens of minutes and an orphan TeV flare have been observed from the blazar Markarian 421 (Mrk 421). The TeV emission from Mrk 421 is believed to be produced by leptonic synchrotron self-Compton (SSC) emission. In this scenario, correlations between the X-ray
and the TeV fluxes are expected, TeV orphan flares are hardly explained and the activity (measured as duty cycle) of the source at TeV energies
is expected to be equal or less than that observed in X-rays if only SSC is considered. To estimate the TeV duty cycle of Mrk 421 and to establish limits on its variability at different time scales, we continuously observed Mrk 421 with the Milagro observatory. Mrk 421 was
detected by Milagro with a statistical significance of 7.1 standard deviations between 2005 September 21 and 2008 March 15. The observed spectrum is consistent with previous observations by VERITAS. We estimate the duty cycle of Mrk 421 for energies above 1 TeV for different hypothesis of the baseline flux and for different flare selections and we compare our results with the X-ray duty cycle estimated by \cite{2009A&A...502..499R}. The robustness of the results is discussed.



\end{abstract}

\keywords{gamma rays: general --- BL Lacertae objects: individual  (Markarian 421)}

\input{ch.intro}


\input{ch.Observation_Mkr_421}

\input{ch.Variability}

\input{ch.dutycycle}

\input{ch.conclusions}

\acknowledgments

The Milagro project has been supported by the National Science Foundation
(under grants PHY-0245234, -0302000, -0400424, -0504201, -0601080,
and ATM-0002744), the US Department of Energy (Office of High-Energy Physics
and Office of Nuclear Physics), Los Alamos National Laboratory, the
University of California, the Institute of Geophysics and Planetary
Physics. This work has been supported by the Consejo Nacional de Ciencia y Tecnolog\'ia (under grants Conacyt 105033 and 103520), Universidad Nacional Aut\'onoma de M\'exico (under grants PAPIIT IN105211 and IN108713) and DGAPA-UNAM.

\addcontentsline{toc}{chapter}{Bibliography}
\begin{large}
\bibliography{bibliography}
\end{large}
\mbox{}

\end{document}

%% file: authorlist.tex
\author{
A.~A.~Abdo,\altaffilmark{\ref{msu},\ref{oed}}
A.~U.~Abeysekara,\altaffilmark{\ref{msu}}
B.~T.~Allen,\altaffilmark{\ref{uci},\ref{cfa}} 
T.~Aune,\altaffilmark{\ref{ucsc},\ref{uca}}
A.~S.~Barber,\altaffilmark{\ref{msu},\ref{utah}}
D.~Berley,\altaffilmark{\ref{umcp}} 
J.~Braun,\altaffilmark{\ref{umcp}}
C.~Chen,\altaffilmark{\ref{uci}}
G.~E.~Christopher,\altaffilmark{\ref{nyu}}
R.~S.~Delay,\altaffilmark{\ref{uci}} 
T.~DeYoung,\altaffilmark{\ref{psu}} 
B.~L.~Dingus,\altaffilmark{\ref{lanl}} 
R.~W.~Ellsworth,\altaffilmark{\ref{georgemason}} 
N.~Fraija,\altaffilmark{\ref{ida}}
M.~M.~Gonz\'alez,\altaffilmark{\ref{ida}} 
J.~A.~Goodman,\altaffilmark{\ref{umcp}} 
E.~Hays,\altaffilmark{\ref{gsfc}}
C.~M.~Hoffman,\altaffilmark{\ref{lanl}}
P.~H.~H\"untemeyer,\altaffilmark{\ref{mich}}
A.~ Imran, \altaffilmark{\ref{lanl}}
B.~E.~Kolterman,\altaffilmark{\ref{nyu}} 
J.~T.~Linnemann,\altaffilmark{\ref{msu}}
A.~Marinelli,\altaffilmark{\ref{idf}}
J.~E.~McEnery,\altaffilmark{\ref{gsfc}}
T.~Morgan,\altaffilmark{\ref{unh}}
A.~I.~Mincer,\altaffilmark{\ref{nyu}} 
P.~Nemethy,\altaffilmark{\ref{nyu}} 
B.~Patricelli,\altaffilmark{\ref{ida}}
J.~Pretz,\altaffilmark{\ref{lanl}}
J.~M.~Ryan,\altaffilmark{\ref{unh}} 
P.~M.~Saz~Parkinson,\altaffilmark{\ref{ucsc}}
M.~Schneider,\altaffilmark{\ref{ucsc}} 
A.~Shoup,\altaffilmark{\ref{osu}} 
G.~Sinnis,\altaffilmark{\ref{lanl}} 
A.~J.~Smith,\altaffilmark{\ref{umcp}} 
V.~Vasileiou,\altaffilmark{\ref{umcp},\ref{france}} 
G.~P.~Walker,\altaffilmark{\ref{lanl},\ref{LasV}} 
D.~A.~Williams\altaffilmark{\ref{ucsc}}
and 
G.~B.~Yodh\altaffilmark{\ref{uci}}} 

\altaffiltext{1}{\label{msu} Department of Physics and Astronomy, Michigan State University, 3245 BioMedical Physical Sciences Building, East Lansing, MI 48824}
\altaffiltext{2}{\label{uci} Department of Physics and Astronomy, University of California, Irvine, CA 92697}
\altaffiltext{3}{\label{cfa} Current address: Harvard-Smithsonian Center for Astrophysics, Cambridge, MA 02138}
\altaffiltext{4}{\label{ucsc} Santa Cruz Institute for Particle Physics, University of California, 1156 High Street, Santa Cruz, CA 95064}
\altaffiltext{5}{\label{umcp} Department of Physics, University of Maryland, College Park, MD 20742}
\altaffiltext{6}{\label{nyu} Department of Physics, New York University, 4 Washington Place, New York, NY 10003}
\altaffiltext{7}{\label{psu} Department of Physics, Pennsylvania State University, University Park, PA 16802}
\altaffiltext{8}{\label{lanl} Group P-23, Los Alamos National Laboratory, P.O. Box 1663, Los Alamos, NM 87545}
\altaffiltext{9}{\label{georgemason} Department of Physics and Astronomy, George Mason University, 4400 University Drive, Fairfax, VA 22030}
\altaffiltext{10}{\label{ida} Instituto de Astronom\'ia, Universidad Nacional Aut\'onoma de M\'exico,
D.F., M\'exico, 04510}
\altaffiltext{11}{\label{gsfc} NASA Goddard Space Flight Center, Greenbelt, MD 20771}
\altaffiltext{12}{\label{utah} Department of Physics, University of Utah, Salt Lake City, UT 84112}
\altaffiltext{13}{\label{idf} Instituto de Fisica, Universidad Nacional Aut\'onoma de M\'exico,
D.F., M\'exico, 04510}
\altaffiltext{14}{\label{unh} Department of Physics, University of New Hampshire, Morse Hall, Durham, NH 03824} 
\altaffiltext{15}{\label{oed} Operational Evaluation Division, Institute for Defense Analyses, 4850 Mark Center Drive, Alexandria, VA 22311-1882}
\altaffiltext{16}{\label{osu} Ohio State University, Lima, OH 45804}
\altaffiltext{17}{\label{uca} Department of Physics and Astronomy, University of California, Los Angeles, CA 90095}
\altaffiltext{18}{\label{mich} Department of Physics, Michigan Technological University, Houghton, MI 49931, USA}
\altaffiltext{19}{\label{france} Laboratoire Univers et Particules de Montpellier, Universit\'e Montpellier 2, CNRS/IN2P3, CC 72, Place Eug\'ene Bataillon, F-34095 Montpellier Cedex 5, France}
\altaffiltext{20}{\label{LasV} National Security Technologies, Las Vegas, NV 89102}

%% file: ch.intro.tex
\section{Introduction}

Mrk 421 is one of the closest (redshift z=0.03; \citealp{1991rc3..book.....D}) and brightest blazars known. 
Due to its low-energy synchrotron peak with $E_{sync} > 0.1$ keV (see, e.g., \citealp{2008ApJ...677..906F}), it is classified as a high-frequency peaked BL Lacertae (HBL) according to the blazar sequence \citep{1995ApJ...444..567P}. Multiwavelenght campaigns, especially in X-rays \citep{2004ApJ...605..662C} and $\gamma$-rays \citep{2010A&A...524A..48T} have shown that Mrk 421 had major outbursts\footnote{A major outburst usually lasts several months and is accompanied by many rapid flares with time scales from tens of minutes to several days, with fluxes varying from a few tenths of Crab up to about ten Crab (see e.g. \citealp{2010A&A...524A..48T}).}. 
Moreover, there is evidence of correlation between simultaneously measured fluxes in the X-ray and TeV energy band \citep{2008ApJ...677..906F}, as expected within the synchrotron self-Compton (SSC) scenario. However, X-ray and VHE emission from Mrk 421 do not always correlate \citep{2006ApJ...641..740R} as was the case of the TeV flare observed without the X-ray counterpart (called ``orphan flare'') by  \cite{2005ApJ...630..130B}. 
Given the limited duty cycle of IACT instruments, one cannot rule out the possibility of lagging counterparts at the other wavelengths.  
Some authors (see e.g. \citealp{2005ApJ...630..186R} and \citealp{2013PhRvD..87j3015S}) have claimed ``orphan flares'' as evidence of hadronic processes taking place in blazars, although non-standard leptonic models (see e.g. \citealp{2006ApJ...651..113K}) can also explain them.

From 2006 to 2008, Mrk 421 was observed by a few instruments. For instance, VERITAS and Whipple observations from 2006 January and 2008 June do not show  significant correlations between the $\gamma-$ray and the optical/radio emission. Moreover, interestingly a $\gamma-$ray flare lasting two days was detected without increased X-ray activity: unfortunately the data in these wavelengths were not exactly contemporaneous to allow the firm conclusion of an orphan TeV flare. MAGIC also reported a flare with rapid flux variability in the time period 2006 April 22-30  \citep{2010A&A...519A..32A}. They also detected a very intense outburst between 2007 December and 2008 June that was studied together with simultaneous data in other wavelengths. They found that it is difficult to describe the SED with the typical variability scale of Mrk 421 within the one zone SSC framework \citep{2012A&A...542A.100A}. ARGO-YBJ observed the flux of Mrk 421 to be correlated with X-ray emission from 2007 November to 2010 February \citep{2010ApJ...714L.208A,2011ApJ...734..110B}. It was pointed out that both the X-ray and $\gamma-$ray spectra harden as the flux increases, favoring the SSC model. 
IACT studies highlight features of specific short activity periods of the source, mainly guided by external or self trigger on high states, that could or could not be attributed to a general behavior of the source. While the sensitivity of Milagro to short duration flares is less than that of IACTs, it is better suited to study long term variabilities and duty cycle, as it operated almost continuously.  Mrk 421 was one of the brightest sources observed by Milagro and was monitored every day for $\sim$ 6 hours. 

In this paper we present the analysis of 3 years of Milagro observations (from 2005 September up to 2008 March) of Mrk 421. We provide upper limits on the flux of a flare, limits on the flux for different time scales, and an estimation of the $\gamma$-ray duty cycle for energies above of 1 TeV of Mrk 421.


%% file: ch.Observation_Mkr_421.tex
\section{Milagro observations: significance map and spectrum of Mrk 421}
\label{sec:skymap}
The Milagro experiment  \citep{2004ApJ...608..680A} was a large water-Cherenkov detector located at  106.68$^o$W longitude, 35.88$^o$N latitude in northern New Mexico, USA at an altitude of 2630 m above sea level that operated from 2000 to 2008. It was designed to detect VHE gamma rays: it was sensitive to extensive air showers (EAS) resulting from primary gamma rays at energies between 100 GeV and 100 TeV \citep{2008ApJ...688.1078A,2008PhRvL.101v1101A}. It had a $\sim$2 sr field  of view and  a $\geq$ 90$\%$ duty cycle that allowed continuous monitoring of the entire overhead sky. The main detector consisted of a central 80 m $\times$ 60 m $\times$ 8 m water reservoir with 723 photomultiplier tubes (PMTs) arranged  in two layers.  The top (air shower) layer (under 1.4 m of water) was equipped with  450 PMTs and the bottom (muon) layer (under 6 m of water) with 273 PMTs.  The air-shower layer was used to reconstruct the direction of the air shower by measuring the relative arrival times of the shower particles across the array. The muon layer was used to discriminate between gamma-ray induced and  hadron-induced air showers. In 2004, a sparse array of 175 ``outriggers'' was added around the central reservoir. The outrigger array covered an area of 40000 m$^2$ and each outrigger was instrumented with a single PMT. This array increased the area of the detector and improved the gamma/hadron separation. The instrument reached its final configuration (physical configuration, analysis procedures and calibration) in 2005 September. This paper only uses data from this last period. 

A detailed description of the Milagro analysis is given in \citet{2012ApJ...750...63A}. Here we summarize the information relevant to this study.

Reconstructed Milagro events (hereafter called events) contain information about the direction (hour angle and declination) of air shower events. From the reconstructed data, sky maps are formed. Sky maps are binned in $0.1^{\circ} \times 0.1^{\circ}$ pixels and contain a signal map with the measured counts on the sky and a background map with the background expectation calculated using the direct integration method described in \citet{2012ApJ...750...63A}. The sky maps are constructed for 9 independent bins of the parameter ${\cal F}$ ($0.2 \leq {\cal F} \leq 2$, in steps of 0.2). This parameter is used to give an estimate of the energy of the primary particle initiating the extensive air-shower and it is defined as
\begin{equation}
{\cal F}=\frac{N_{AS }}{N_{AS}^{live}}+\frac{N_{OR}}{N_{OR}^{live}},
\end{equation}
where $N_{AS (OR)}/N_{AS (OR)}^{live}$ is the ratio between the number of PMTs in the air-shower layer (AS) / outriggers (OR)  detecting the event and the number of functional PMTs in the air-shower layer (AS) / outriggers (OR) at that time. More energetic showers contain more particles, cover a larger area, and so fire more PMTs: a higher value of the parameter ${\cal F}$ is then obtained. The dependence of the parameter ${\cal F}$ with the energy of the primary particle is shown in Fig. 5 of \citet{2012ApJ...750...63A}. 

To maximize the statistical significance when searching for sources, a weighted analysis technique is used \citep{2007PhDT........19A,2012ApJ...750...63A}. A weight is applied to all events, in both signal and background maps: gamma-like events are given  higher weights than cosmic-ray like events.  The values of these weights also depend on the parameter ${\cal F}$ to account for the angular resolution of the detector, which is a function of the size of the event and of the parameter ${\cal F}$. The angular resolution ranges from 1.2$^{\circ}$ for small values of ${\cal F}$ to 0.35$^{\circ}$ for large values of ${\cal F}$ \citep{2012ApJ...750...63A}. To calculate the weights for each ${\cal F}$ bin, the standard Milagro optimization hypothesis for the spectrum of an extragalactic source, consisting of a power-law with exponential cut-off at an energy of 5 TeV and photon index of 2.0, is assumed. This spectrum roughly includes the EBL absorption. The results are not strongly dependent on the  exact shape of the assumed spectrum.  
For each ${\cal F}$ bin, the gamma-ray like excess with respect to the background is calculated as the difference between signal weighted events and background weighted events. This results in nine excess sky maps, one for each ${\cal F}$ bin. Finally, all nine excess sky maps are added into a final excess sky map. The sky map of the statistical significance is obtained by using  Equation 4 of \cite{2012ApJ...750...63A}.

\begin{figure}[h!]
\begin{center}
\includegraphics[width=100mm]{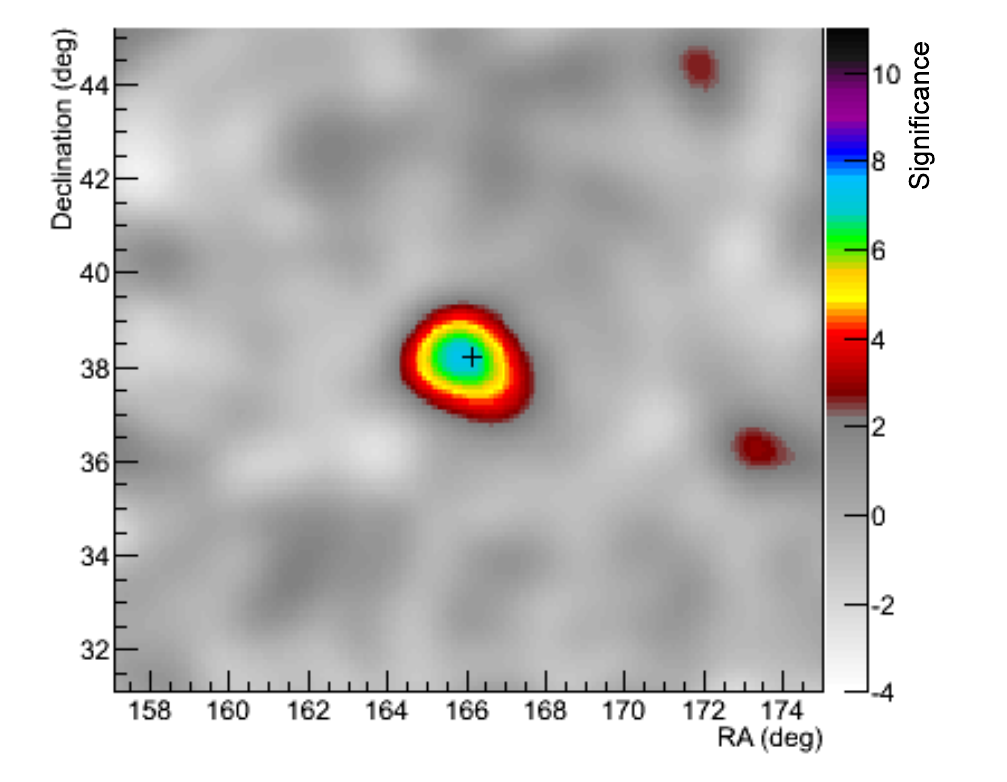}
\caption{Sky map of the statistical significance in the region of Mrk 421. The significance at the Mrk 421 location (black cross) is 7.1 standard deviations.}\label{fig:skymap}
\end{center}
\end{figure}

Mrk 421 was observed with a significance of 7.1 standard deviations for a period of 906 days (828 integrated days after data quality cuts) from 2005 September 21 to 2008 March 15. The median energy of the detected gamma rays is 1.7 TeV (under the spectral optimization hypothesis given above). The final map of the statistical significance of the excesses in the region around Mrk 421 is shown in Figure~\ref{fig:skymap}. 

VERITAS has measured the spectrum of Mrk 421 in different flux states (classified by level of intensity, from ``very low''  to ``very high'', \citealp{2011ApJ...738...25A}). In all cases, the energy spectrum cuts off below 10 TeV, and an exponential cut-off at 4 TeV is most typical. A fit to the energy spectrum with Milagro data, using the same approach employed to measure the Crab spectrum \citep{2012ApJ...750...63A}, provides a limited constraint to the spectrum because the emission from Mrk 421 is concentrated at the lowest energy range of Milagro's sensitivity. Nevertheless, we can use the Milagro data to test a specific spectral assumption for consistency. As with the Crab measurement, we determine the ${\cal F}$ distribution from the source and generate an expected ${\cal F}$ distribution for several assumed spectra, determining a $\chi^2$ to characterize the agreement between that hypothesis and the data. We find that the VERITAS ``low" spectrum is most consistent with Milagro data with a $\chi^2$ of 12.7 and 9 degrees of freedom. The VERITAS ``very low" spectrum is marginally inconsistent with the 3-year integrated average, with a $\chi^2$ of 31.1 and 9 degrees of freedom. The ``mid" spectrum is inconsistent with a $\chi^2$ of 124.1 and 9 degrees of freedom, primarily because of the normalization, rather than the spectral shape. Fixing the low-energy spectral index at 2.3 that has been measured for Mrk 421 by VERITAS at low-TeV energies, we find an exponential cut-off energy between 2.2 TeV and 5.6 TeV at one standard deviation of confidence, consistent with VERITAS measurements.

%% file: ch.Variability.tex
\section{Variability}
\label{sec:variability}

The light curve of Mrk 421 is obtained by converting the measured weighted event excesses (hereafter called weighted excesses) into fluxes through the calculation, with Monte Carlo simulations, of the expected weighted excesses for an assumed Mrk 421 spectrum. The assumed spectrum is taken with an index of 2.3, a cut-off energy of 4 TeV and a normalization of $\rm 0.46\times 10^{-10} cm^{-2} s^{-1} TeV ^{-1}$.  The spectral index is consistent with the VERITAS ``low state'', the energy cut-off of 4 TeV is the most typical value for Mrk 421 \citep{2001ApJ...560L..45K,2002A&A...393...89A,2005A&A...437...95A,2008ApJ...679L..13K,2011ApJ...738...25A} and, the normalization is obtained by fitting the spectrum as described in the previous section. Thus, the measured weighted excesses are divided by the expected weighted excesses and then multiplied for the integrated flux of the assumed spectrum for energies above 1 TeV ($\rm 0.20 \times 10^{-10} cm^{-2} s^{-1}$). 

\begin{figure}[h!]
\includegraphics[width=180mm]{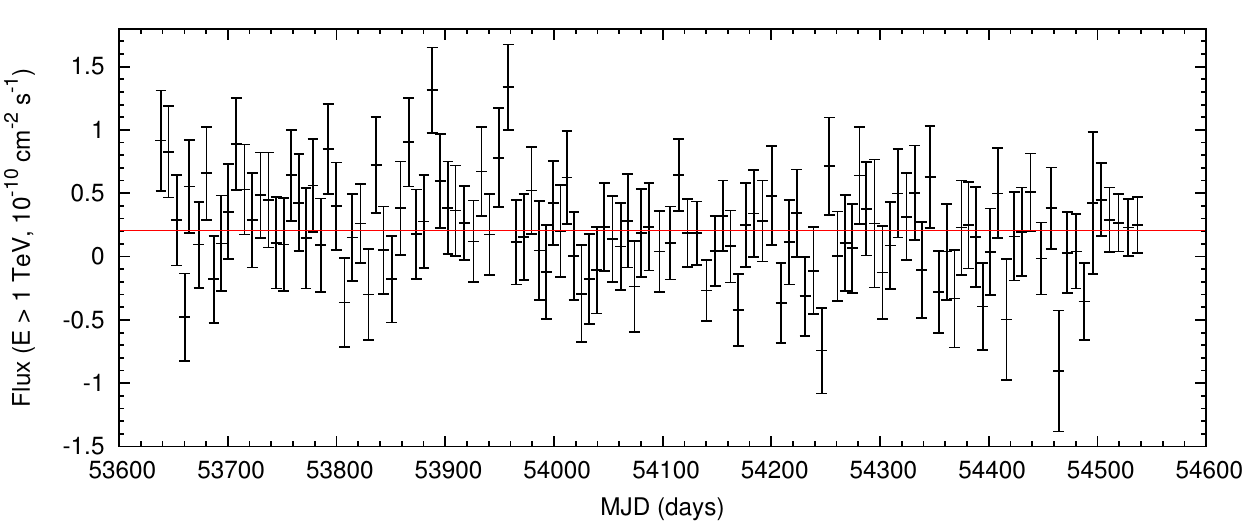}
\caption{Light curve of Mrk 421 (black points) for energies above 1 TeV; the red solid line represents the average value of the flux: $\bar F_{\rm Milagro}$=($0.205 \,\pm 0.030$) $\times 10^{-10}\,\rm{cm^{-2}\,s^{-1}}$. Each bin represents $\sim$1 week of data.}\label{fig:LC} 
\end{figure}

The light curve (LC) obtained with Milagro for energies above 1 TeV  is shown in Figure \ref{fig:LC}. Milagro data were recorded on tapes with each tape containing data collected over a time interval that, on average, is about 1 week. Each time bin in the light curve corresponds to data recorded in one tape. 

If we assume a constant flux from Mrk 421, we obtain an average flux  for energies above 1 TeV of $\bar F_{\rm Milagro}$=($0.205 \,\pm 0.030$) $\times 10^{-10}\,\rm{cm^{-2}\,s^{-1}}$. This is consistent with the spectrum used to calculate the expected weighted excesses. The $\chi^2$ is 134 for 122 degrees of freedom, which gives a $\chi^2$ probability of 21 \%, indicating that the Mrk 421 flux, measured by Milagro, is consistent with being constant during the 3-year monitoring period. This average flux corresponds to (0.85 $\pm$ 0.13) Crab, using the Crab flux as measured by Milagro \citep{2012ApJ...750...63A} for energies above 1 TeV. 

\begin{figure}[h!]
\includegraphics[width=180mm]{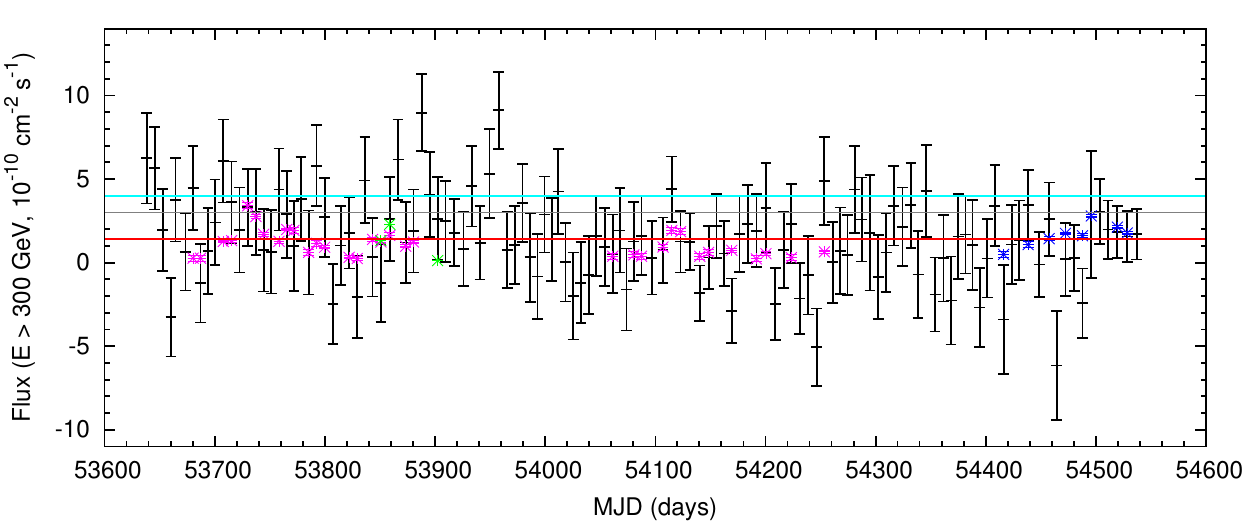}
\caption{Light curve of Mrk 421 for energies above 300 GeV. In black: Milagro data; in magenta and in blue: data by Whipple and VERITAS respectively \citep{2011ApJ...738...25A}; in green: MAGIC fluxes, calculated as the integral above 300 GeV of the measured spectra \citep{2010A&A...519A..32A}. Data from IACT observatories have been combined to match the Milagro time-binning. Also shown are in red the Milagro average flux (1.4$\times 10^{-10}\,\rm{cm^{-2}\,s^{-1}}$), in grey and cyan the fluxes corresponding to the VERITAS mid and lowest high state (``high state A'') of Mrk 421 (3.1$\times 10^{-10}\,\rm{cm^{-2}\,s^{-1}}$ and 4.1$\times 10^{-10}\,\rm{cm^{-2}\,s^{-1}}$ respectively, calculated as the integral above 300 GeV of the corresponding best fit spectrum with a fixed cut-off energy of 4 TeV, see \citealp{2011ApJ...738...25A}).}\label{fig:LC300GeV}
\end{figure}

We also compute the LC of Mrk 421 for energies above 300 GeV, shown in Figure \ref{fig:LC300GeV}, to make a direct comparison with other VHE observations (see \citealp{2010A&A...519A..32A,2011ApJ...738...25A}). The data from the other instruments have been combined\footnote{The combined average flux has been calculated only by considering the days with reported fluxes and assuming the flux to be the same for the whole week.} to match the Milagro binning. 

As mentioned in the previous section, the spectrum observed by Milagro is consistent with the spectrum in the low state observed by VERITAS and with being constant over time. Therefore, either there are many bright flares that last a much shorter time than a week
, or there are only a few very bright flares such that the average flux over years is still consistent with a low state. The fluxes corresponding to the mid and high states reported by VERITAS are also shown in Figure \ref{fig:LC300GeV}. IACT observations of Mrk 421 during this period indicate that it was not in a high state on week time scales and only for one week it was just above the mid state. Clearly, this statement does not stand for much shorter time scales than about a week as one can notice from the original IACTs LCs \citep{2010A&A...519A..32A,2011ApJ...738...25A}. 

\begin{figure}[h!]
\includegraphics[width=180mm]{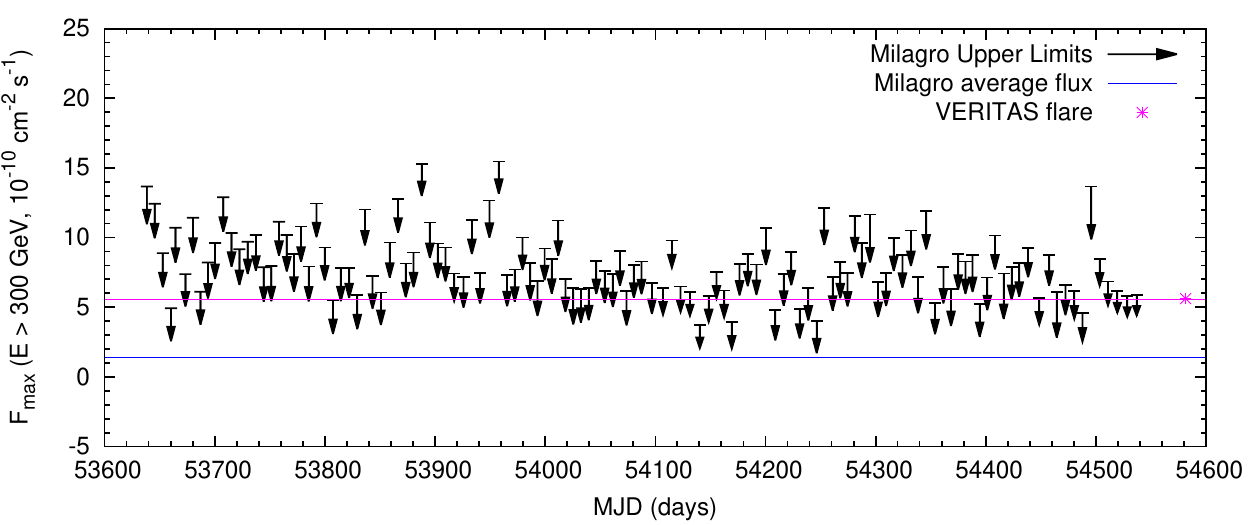}
\caption{ Black arrows: maximum averaged flux in a week time period for a flare in order to have not been detected at a 99.7 $\%$ C.L.; the length of the downward arrow for each point is the equal to the size of the corresponding error bar on the flux. 
Blue solid line: average Milagro flux above 300 GeV. Magenta star: observation by VERITAS corresponding to a flaring state, see  \citet{2011ApJ...738...25A}; VERITAS data have been combined to have a weekly time-binning. The magenta line marks the flux level of the VERITAS flare along the whole Milagro observation period.}\label{fig:MaxFlare}
\end{figure}

Some of the Milagro measurements correspond to fluxes consistent with the high state. However, this is just a result of the statistical fluctuations associated with the large error bars of each measurement. Thus, it cannot be concluded from Milagro data that Mrk 421 was observed in a high state for those bins. In fact, all measurements are within 3 standard deviations of the Milagro average flux except the two at 53888 and 53958 MJD. These bins are above the Milagro average flux at significance of 3.28 and 3.34 $\sigma$, respectively, but only 1.54 and 1.64 $\sigma$ after correcting for trials. Therefore there is no significant evidence for flares in Milagro data. We can calculate the maximum average flux, $F_{\rm max}$, in a week time period for a flare not to have been detected at 99.7 \% confidence level (C.L.) using the method of Helene \citep{1983NIMPR.212..319H}. In Figure~\ref{fig:MaxFlare} we show the flux measurements given in Figure~\ref{fig:LC300GeV} converted to upper limits (99.7 \% C.L.) on the flux above 300 GeV. The length of the downward arrow for each point is the equal to the size of the corresponding error bar of each Milagro flux measurement. For comparison, we show the Milagro average flux and, the flare observed by VERITAS on 2008 May with a maximum flux of $\sim 12  \times 10^{-10}$ cm$^{-2}$ s$^{-1}$, as reported in  \cite{2011ApJ...738...25A}. If we combine$^2$ the VERITAS data to have a weekly time-binning as Milagro, the resulting average value is $\sim 5.6 \times 10^{-10}$ cm$^{-2}$ s$^{-1}$.

\begin{figure}[h!]
\includegraphics[width=180mm]{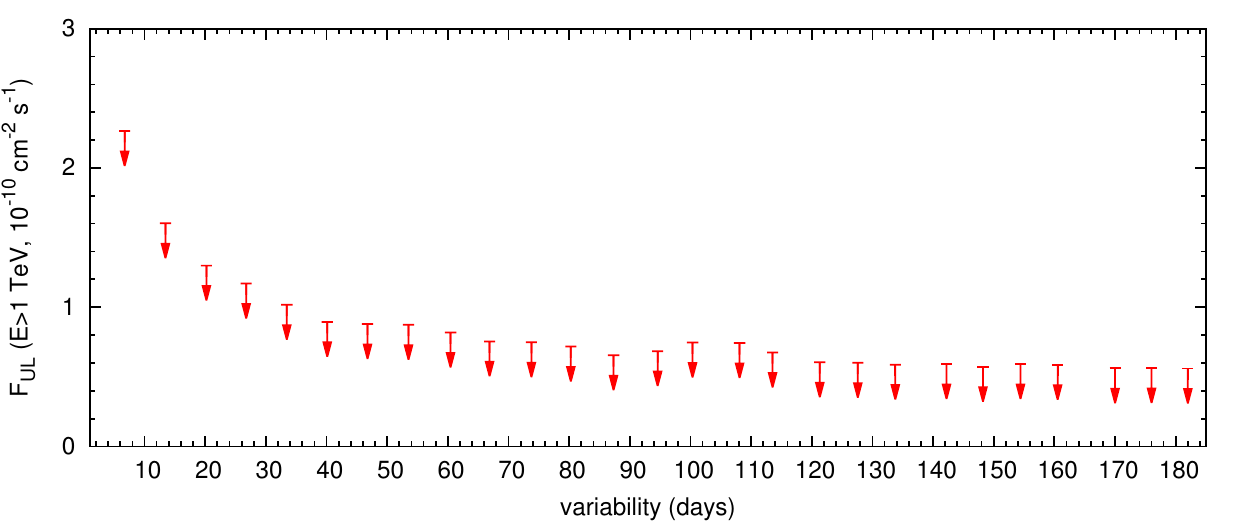}
\caption{Upper limits on the flux as a function of the flare duration.}\label{fig:UL} 
\end{figure}

We have also calculated the largest value of the maximum averaged flux ($F_{\rm max}$) for flares of different durations. We have binned the data in several intervals from one week to six months to account for different variability time scales, as outbursts have been observed to last up to several months (see e.g. \citealp{2010A&A...524A..48T}). We have calculated the flux upper limits above an energy of 1 TeV. As observed in Figure~\ref{fig:UL}, the values of the flux upper limits vary from \mbox{2.26 $\times$ 10$^{-10}$ cm$^{-2}$ s$^{-1}$} to 0.56 $\times$  10$^{-10}$ cm$^{-2}$ s$^{-1}$ for a variability time scale of a week to six months, respectively.

HEGRA observed a flare that lasted about three months \citep{2003A&A...410..813A}. The maximum flux of the flare observed in the night of 2001 April 1st, was \mbox{2.5 $\rm \times 10^{-10} cm^{-2} s^{-1}$}. Thus, considering the Milagro upper limits for one (2.26 $\rm \times 10^{-10} cm^{-2} s^{-1}$) and two (1.6 $\rm \times 10^{-10} cm^{-2} s^{-1}$) weeks, a flare with the maximum flux observed by HEGRA could not have lasted longer than one week without having been detected by Milagro. 

%% file: ch.dutycycle.tex
\section {Duty cycle}\label{sec:dutycycle}
As previously mentioned, blazars are highly variable sources 
in short time scales. The lowest steady flux level is called the baseline state. The level of activity of a source can be measured as the percentage of time that the source spends in flaring states, also called duty cycle, given by,

\begin{equation}\label{eq:DC}
{\rm duty\,cycle}\equiv DC=\frac{\sum_i t_i}{T_{\rm obs}},
\end{equation}

where $t_i$ is the time that the source spends in the $i$-flaring state, with $i$ running over all the flaring states in the observation period $T_{\rm obs}$.

To calculate the duty cycle a flare flux threshold must be established to distinguish flaring states. For example, \cite{2004ApJ...601..151K} estimated the X-ray duty cycle of several blazars (including Mrk 421) as the fraction of time during which the flux exceeds the flux threshold equivalent to 150 \% of the time averaged flux. The same flare flux threshold was used by \cite{2008MNRAS.385..119W}, with the additional condition that the 50 \% deviation from the time average flux (considered by \citealp{2004ApJ...601..151K}) was required to be significant at the 3$\sigma$ level.  \cite{2007JPhCS..60..318T} estimated the TeV duty cycle of Mrk 421 by using arbitrary flare flux thresholds of 1, 2, 3, 4 and 5 Crab.
Finally, \cite{2009A&A...502..499R} estimated the baseline flux $R_{char}$ and the associated error $\sigma_{char}$ for Mrk 421 in X-rays and then calculated the X-ray duty cycle by considering a flare flux threshold equal to $R_{N\sigma}=R_{char}+N \sigma_{char}$, for $N$=1, 3. In other words, flaring states are those whose fluxes are 1 and 3 standard deviation above the baseline flux. 

Choosing a flare flux threshold in terms of the time average flux does not allow a direct comparison of duty cycles between sources and between different energy bands to be made because  the time average flux is influenced by the level of activity. In the case of a highly active source, the time average flux, and consequently its flare flux threshold,  would be much higher than the baseline flux, so the duty cycle only refers to the highest flux states. On the contrary, for a less active source the time average flux is  close to the baseline flux, so the duty cycle refers to almost all the flaring states. Since the duty cycle of the active source includes only the highest flux flaring states, it is possible to obtain a duty cycle value smaller than the one for the less active source and erroneously conclude that the latest source is more active than the former one. A similar situation happens when comparing duty cycle in different energy bands for a given source. The case of an arbitrary choice of the flare flux threshold will also lead to wrong conclusions, since the same threshold selects higher (compared to the baseline flux) flaring states in the less active source. Alternatively, the flare flux thresholds defined by \cite{2009A&A...502..499R} selects flaring states with fluxes equally significant (in terms of standard deviations) when compared with the baseline flux, independent of the source and of the energy band considered. 
Therefore, we have adopted the flare flux threshold definitions proposed by  \cite{2009A&A...502..499R}, in order to get an estimate of the TeV duty cycle of Mrk 421 to be directly comparable with the X-ray duty cycle.

The duty cycle definition (Equation \ref{eq:DC}) cannot be used directly because of the lack of a complete and systematically observed set of TeV flux states over a period of years, as in the case of X-ray energy band. Instead, we take advantage of the time average flux observed by Milagro as follows:  
the total fluence observed by Milagro, $\bar F_{\rm Milagro}\times T_{\rm Milagro}$ (see Section \ref{sec:variability}),  where $T_{\rm Milagro}$ is the Milagro observation period, can be decomposed into the fluence from the baseline state and the fluence from flaring states, i.e. 
\begin{equation}\label{eq:fluence}
\bar F_{\rm Milagro} \times T_{\rm Milagro}= F_{\rm baseline}\times T_{\rm baseline}+\sum_i F_{{\rm flare},i}\, t_i,
\end{equation}

where $F_{\rm baseline}$ is the baseline flux, $T_{\rm baseline}$ is the time that the source spent in the baseline state,  $F_{{\rm flare},i}$ is the average flux of the $i$ flaring state in a time scale $t_i$, with $i$ running over all the flares in the Milagro  observation period.

As discussed in Sections 2 and 3,   $\bar F_{\rm Milagro}$ is constant over $T_{\rm Milagro}$ and consistent with the ``low" state (not the lowest)  observed by VERITAS. Thus, we could infer that the contribution from the flaring states is small. However this is not necessarily correct as it depends on the value of the baseline flux. The contribution of the flaring states to the total fluence alone does not determine the activity of the source, since the same value could be obtained by considering many long-duration low-flux flaring states or a few short-duration high-flux flaring states. 

$T_{\rm baseline}$ is equal to $T_{\rm Milagro}-T_{\rm flare}$, where $T_{\rm flare}=\sum_i t_i$ is the total time that the source spends in flaring states. If the second term in the right side of Equation \ref{eq:fluence} is rewritten as  $T_{\rm flare} \times<F_{\rm flare}>$, where $<F_{\rm flare}>$ is the average flux of flaring states, we can solve Equation \ref{eq:fluence} for $T_{\rm flare}$ 
and then Equation \ref{eq:DC} becomes,

\begin{equation}\label{eq:DC2}
DC=\frac{\left( \bar F_{\rm Milagro}-F_{\rm baseline}\right)}{<F_{\rm flare}>-F_{\rm baseline}} 
\end{equation}

(where we have used $T_{\rm Milagro}=T_{\rm obs}$).


The estimation of $<F_{\rm flare}>$ is obtained from the distribution of day-wise fluxes of Mrk 421 collected  by \citet{2010A&A...524A..48T} from several VHE experiments (HEGRA, HESS, MAGIC, CAT and Whipple/VERITAS) from 1992 to 2009. \citet{2010A&A...524A..48T} combined the day-wise light curves from different experiments by converting the measured flux values to flux values, $F$, in units of the Crab Nebula flux and normalizing to a common energy threshold of 1 TeV. This was done by using the energy spectrum of the Crab Nebula as measured by each experiment. The resulting distribution is well described by a function, $g(F)$, which is the sum of: 1) a Gaussian, $g_{\rm G}(F)$, whose mean equal to (0.3285$\pm 0.0249$)  Crab ($\sim$ 0.33 Crab) represents the upper limit on $F_{\rm baseline}$ and 2) a {\rm log}-normal function $g_{\rm ln}(F)$ that describes the flaring states (see Figure 3 in \citealp{2010A&A...524A..48T}): 

\begin{equation}\label{eq:gauss}
g_{\rm G}(F)=\frac{N_{\rm G}}{\sigma_{\rm G}\,\sqrt{2 \pi}}\,\exp\left[-\frac{1}{2}\left(\frac{F-\mu_{\rm G}}{\sigma_{\rm G}}\right)^2 \right]
\end{equation}
and
\begin{equation}
g_{\rm ln}(F)=\frac{N_{\rm ln}}{F\,\sigma_{\rm ln}\,\sqrt{2 \pi}} \exp\left[-\frac{({\rm log}(F)-\mu_{\rm ln})^2}{2 \sigma_{\rm ln}^2} \right].
\end{equation}

Therefore, we have extrapolated, when needed, the function $g_{\rm ln}(F)$ down to the flare flux threshold $F_{\rm thr}$ and calculated the average flare flux as follows: 

\begin{equation}\label{eq:ffm}
<F_{\rm flare}>=\frac{\int_{F_{\rm thr}}^{F_{\rm lim}} F\,g_{\rm ln}(F)\,dF}{\int_{F_{\rm thr}}^{F_{\rm lim}} g_{\rm ln}(F)\,dF}
\end{equation}

where $F_{\rm lim}$= 10 Crab is the maximum flux observed in the distribution by \citet{2010A&A...524A..48T}. $<F_{\rm flare}>$ depends on the value of  $F_{\rm thr}$. For instance, $<F_{\rm flare}>$ is 1.67, 1.84 and 2.64 Crab for $F_{\rm thr}$ of 0, 0.33 and 1 Crab, respectively.
 

Given Equations \ref{eq:DC2} and \ref{eq:ffm}, we have calculated the duty cycle for several different assumptions of $F_{\rm baseline}$ and for two different flare flux thresholds: $F_{\rm thr}=F_{\rm baseline}+ N\times \sigma_{\rm G}$, with $N$=1, 3 (with $\sigma_G$ defined in Eq. \ref{eq:gauss}). We have also estimated its uncertainties as a function of  the errors associated with $g_{\rm ln}(F)$,  $F_{\rm lim}$ and  $\bar F_{\rm Milagro}$.

 

The uncertainty in the TeV duty cycle related to the errors in the parameters of $g_{\rm ln}(F)$  has been estimated to be of 4 \% \citep{RICAPProc}. The extrapolation of $g_{\rm ln}(F)$ for $F_{\rm lim}$ above 10 Crab is not trivial since it depends on several factors such as, e.g., the total available energy of the source and the capability to maintain a high flux for a time equal to the duration of the flux states  considered by \citet{2010A&A...524A..48T}. Nevertheless, we find that changing $F_{\rm lim}$ from 10 Crab to 15 Crab lowers the calculated duty cycle by between 6 and 8  \% depending on the baseline flux \citep{IcrcProc}. In the following analysis we do not make further assumptions on $F_{\rm lim}$ and we only consider the case $F_{\rm lim}=10$ Crab. The uncertainty on the duty cycle values only considers the error associated with $\bar F_{\rm Milagro}$.



\begin{figure}[h!]
\begin{center}
\includegraphics[width=83mm]{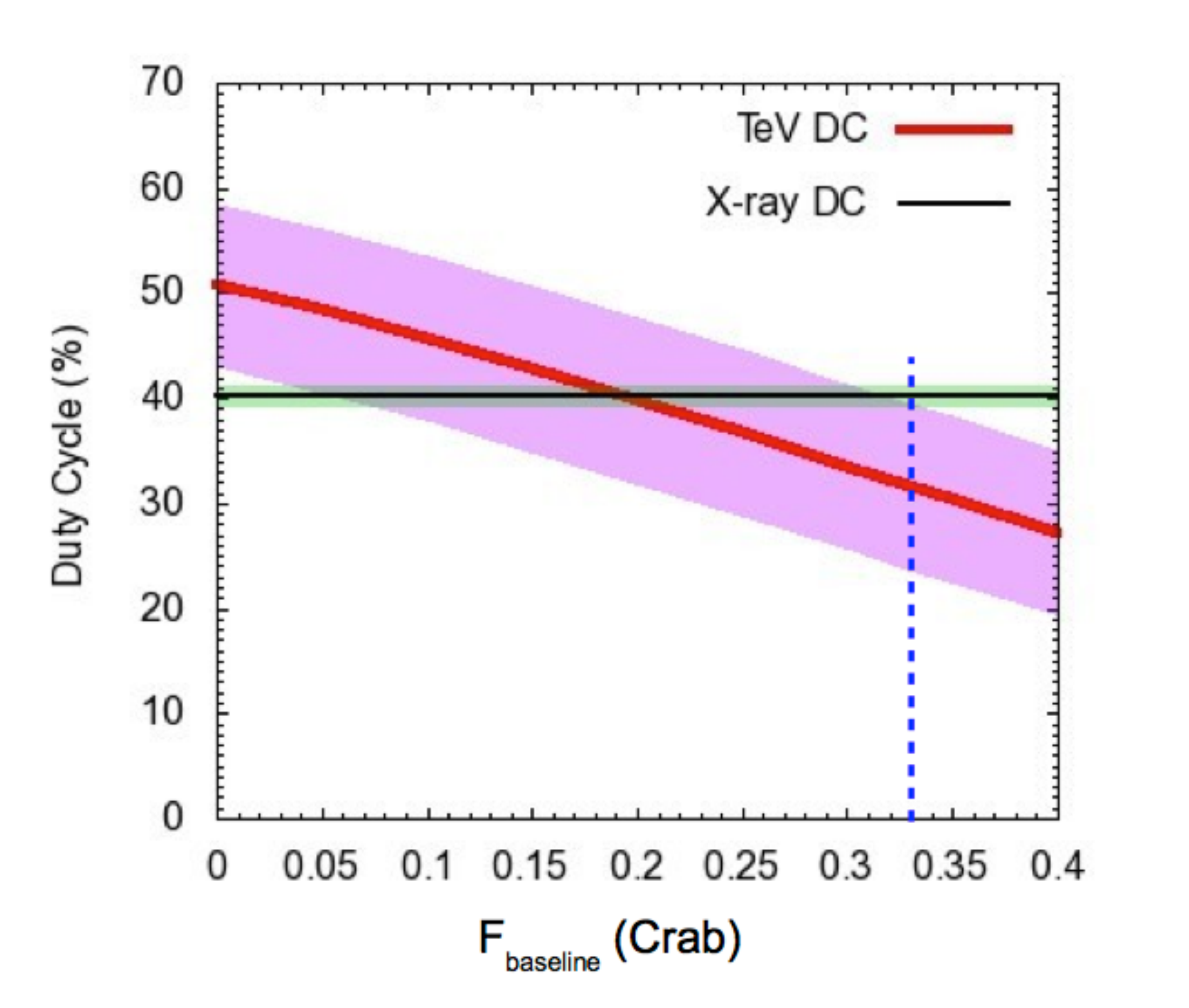}\includegraphics[width=80mm]{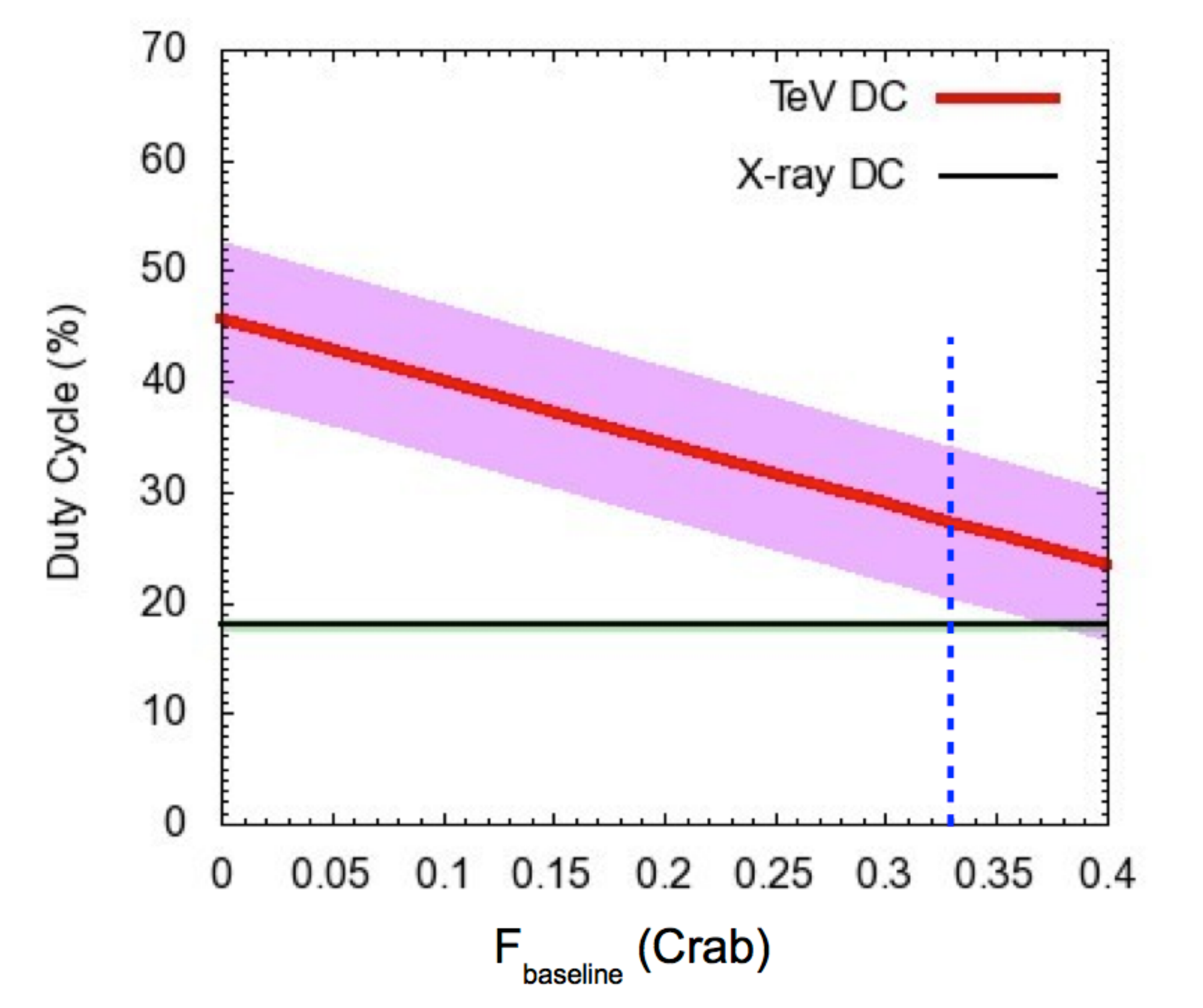}
\caption{In red: TeV duty cycle calculated by considering as flares all the states having a flux above 1 TeV greater than $F_{\rm baseline}$+N$\sigma_{\rm G}$ (left: N=1, right: N=3); the shadowed pink area represents the error associated to the uncertainty on $\bar f$. In black: X-ray duty cycle, calculated by considering as flares all the states having a flux greater that $R_{\rm char}$+N$\sigma_{char}$ (left: N=1, right: N=3,  \citealp{2009A&A...502..499R}); the shadowed green area represents its error. The dashed blue line marks the duty cycle values for $F_{\rm baseline}$=0.33 Crab.}
\label{fig:DC}
\end{center}
\end{figure}

The value of duty cycle, in the case of $N=1$ (shown in the left panel of Figure ~\ref{fig:DC}) ranges from (51 $\pm$ 8) \% to (32 $\pm$ 8) \% for $F_{\rm baseline}$=0 Crab and 0.33 Crab, respectively, while for the case of $N=3$ (shown in the right panel of Figure ~\ref{fig:DC}) it ranges from $(46 \pm 7)\,\%$ to $(27 \pm 7) \,\%$ for $F_{\rm baseline}$=0 Crab and 0.33 Crab, respectively. 

For comparison, the X-ray duty cycle values determined by \cite{2009A&A...502..499R} are represented by black lines in Figure \ref{fig:DC}. For the case of $N=1$, the X-ray duty cycle equal to (40.3 $\pm$ 1.0) \%  is consistent, within the error bars,  with the TeV duty cycle almost independent of the value of $F_{\rm baseline}$. This result could be explained if the X-ray and the TeV activity of the source are tightly coupled. However, it should be considered that the X-ray duty cycle may be overestimated since  fluctuations in the X-ray baseline flux have not been discriminated from the flaring states, contrary to what we have done for the TeV duty cycle calculation by using $g_{\rm ln}(F)$ instead of $g(F)$.
The X-ray duty cycle, for $N=3$, is equal to (18.1 $\pm$ 0.5) \%, slightly lower than the TeV duty cycle independent of the assumed value of $F_{\rm baseline}$; however, the uncertainty in the TeV duty cycle is too large to claim a higher activity in TeV than in X-rays. The TeV duty cycle becomes consistent with the X-ray duty cycle for $F_{\rm lim} >$ 18 Crab. We should notice that this result is sensitive to other possible emission mechanisms besides the SSC and that the TeV duty cycle refers to the 3-years of Milagro monitoring, while the X-ray duty cycle refers to a period of more than 10 years: to do a more direct comparison between the two duty cycle, they should be calculated with data collected over the same period of time. 


For completeness, we also considered a flare flux threshold as given by  \cite{2004ApJ...601..151K}. In this case, the TeV duty cycle cannot be calculated directly from Eq.  \ref{eq:DC2} since 
$F_{\rm thr}=F_{\rm thr}^{K}$=1.275 Crab corresponds to a flux value of $\sim$ 9 standard deviations above the baseline flux (considering $F_{\rm baseline}=0.33$ Crab). 
Thus we cannot assume, as done in the cases $N=1,3$, that the states with flux lower than $F_{\rm thr}$ are only fluctuations of the baseline flux. Instead, these states should be counted in the total time as well as their flux into the total fluence. 


From Eq. \ref{eq:DC}, the ratio between the duty cycle calculated with a flare flux threshold $F_{\rm thr}^K$ and any other flare flux threshold $F_{\rm thr}$ is given by
\begin{equation}\label{eq:ratioDC}
\frac{DC(F_{\rm thr}^K)}{DC(F_{\rm thr})}=\frac{T_{\rm flare}(F_{\rm thr}^K)}{T_{\rm flare} (F_{\rm thr})},\end{equation}
where $T_{\rm flare}(F_{\rm thr}^K)$ and $T_{\rm flare}(F_{\rm thr})$ are the total time spent in flaring states with fluxes above  $F_{\rm thr}^K$ and $F_{\rm thr}$ respectively and are proportional to the number of the corresponding flaring states. Thus Eq. \ref{eq:ratioDC} can be rewritten as: 
\begin{equation}\label{eq:DCalt}
DC(F_{\rm thr}^K)=\frac{DC(F_{\rm thr})} {\int_{F_{\rm thr}}^{F_{\rm lim}}g(F) dF} \times \int_{F_{\rm thr}^K}^{F_{\rm lim} }g(F) dF
\end{equation}

Note that the quantity $\frac{DC(F_{\rm thr})}{\int_{F_{\rm thr}}^{F_{\rm lim}}g(F) dF}$ is independent of $F_{\rm thr}$, so we can calculate $DC(F_{\rm thr}^K)$ using any of the previous estimated values of duty cycle. We then find that the TeV duty cycle calculated with a flare flux threshold $F_{\rm thr}^K$ ranges from (22$^{+4}_{-3}$) \% to (17$^{+4}_{-4}$) \% for $F_{\rm baseline}$=0 and 0.33 Crab, respectively. These values must not be directly compared  with the X-ray duty cycle obtained by \cite{2004ApJ...601..151K} as the latter corresponds only to flaring states with fluxes above 2 standard deviations from the X-ray baseline flux\footnote{In \cite{2004ApJ...601..151K} the numerical value of the flare flux threshold is not reported and cannot be directly estimated, as the average X-ray flux used is not given. To get an estimate of this threshold  we considered the X-ray average flux estimated by \cite{2008MNRAS.385..119W}, equal to 0.86 cts/s. With this value the flare flux threshold corresponding to the definition of flares of \cite{2004ApJ...601..151K} is 1.29 cts/s. This value corresponds to $\sim R_{char}+2 \sigma_{char}$, with $R_{char}$ and $\sigma_{char}$ given by \citealp{2009A&A...502..499R} (0.5 and 0.4 counts/s respectively).} instead of 9 standard deviations, as considered at TeV energies. The fact that the time average TeV flux (as measured by Milagro) is much higher than the time average X-ray flux (see footnote 3), when compared with the corresponding baseline fluxes (for TeV with the upper limit of 0.33 Crab given by \citealp{2010A&A...524A..48T}, for X-rays see footnote 3), may be an indication that the source is more active in TeV energies than in X-rays. However, simultaneous observations are needed to make a firm statement.

Finally, using Eq. \ref{eq:DCalt} we have obtained a TeV duty cycle for a flare flux threshold of 1 Crab  that  ranges from 27$^{+4}_{-5}$ \% to 21$^{+5}_{-5}$ \% for $F_{\rm baseline}$=0 Crab and 0.33 Crab, respectively. These values are lower than (43 $\pm$ 13) \%, the value calculated by \citet{2007JPhCS..60..318T}, for the ratio of the time in which the source was observed in a flaring state and the total observation time of the telescopes (IACTs).  This is not surprising, as their duty cycle may be overestimated because of the observational bias of IACTs to continue observations when the source is in a high state, leading to an underestimation of the number of observations in the baseline state. 

%% file: ch.conclusions.tex
\section{Conclusions}
\label{sec:conclusions}
We have presented results from a 3-year long term continuous monitoring of the BL Lac Mrk 421 with the Milagro water Cherenkov observatory sensitive to  gamma rays between 100 GeV and 100 TeV. Mrk 421 was detected with a statistical significance of 7.1 standard deviations over the period from 2005 September 21 to 2008 March 15. The Milagro measured spectrum is consistent with the VERITAS ``low state''.  Fixing the spectral index at 2.3 as measured by VERITAS, we found an exponential energy cut-off between 2.2 and 5.6 TeV, also consistent with the VERITAS measurements. We have also found that the Mrk 421 average flux for energies above 1 TeV equals ($0.205 \,\pm 0.030$) $\times 10^{-10}\,\rm{cm^{-2}\,s^{-1}}$, consistent with being constant along the Milagro observation period.


We have found no evidence for flares in the Milagro data. Therefore we have established flare flux upper limits for energies above 300 GeV for a time scale of $\sim$  1 week as a function of time. In addition, we have calculated upper limits on the flare flux for energies above 1 TeV for time scales from one week up to six months, finding that they vary from \mbox{2.26 $\times$ 10$^{-10}$ cm$^{-2}$ s$^{-1}$} to 0.56 $\times$  10$^{-10}$ cm$^{-2}$ s$^{-1}$, respectively.

Such long-term continuous monitoring has allowed us to calculate the $\gamma$-ray duty cycle of Mrk 421 for flaring states with different flare flux thresholds. We have discussed different procedures to define the flare flux threshold and justified the reasons to adopt the definition given by  \cite{2009A&A...502..499R} in our analysis. Two cases are presented in detail: flare flux threshold of 1 and 3 standard deviations above the baseline flux. 
We have compared the corresponding results (see Figure \ref{fig:DC}) with the X-ray duty cycle estimated by \cite{2009A&A...502..499R}. We find that the TeV duty cycle is consistent with the X-ray duty cycle and therefore with the SSC  emission mechanism, although it is sensitive to alternative emission processes. More observations and further studies, for instance of the expected correlation between the activity at TeV energies and X-rays, are required to reduce the uncertainties in the quantities involved in the duty cycle calculation and to obtain a conclusive result on the emission mechanisms involved.

The High Altitude Water Cherenkov detector (HAWC), the successor of Milagro, will be able to produce a more accurate analysis of the TeV emission from Mrk 421, with its greater sensitivity (10-15 times better than Milagro). In particular, with HAWC it will be possible to determine with greater accuracy the average flux, as well as the distribution of flux states of Mrk 421, allowing a more precise estimation of the TeV duty cycle.

%% file: text.bbl
\newcommand{\noop}[1]{}
\begin{thebibliography}{}
\expandafter\ifx\csname natexlab\endcsname\relax\def\natexlab#1{#1}\fi

\bibitem[{{Abdo}(2007)}]{2007PhDT........19A}
{Abdo}, A.~A. 2007, PhD thesis, Michigan State University

\bibitem[{{Abdo} {et~al.}(2008{\natexlab{a}}){Abdo}, {Allen}, {Aune}, {Berley},
  {Blaufuss}, {Casanova}, {Chen}, {Dingus}, {Ellsworth}, {Fleysher},
  {Fleysher}, {Gonzalez}, {Goodman}, {Hoffman}, {H{\"u}ntemeyer}, {Kolterman},
  {Lansdell}, {Linnemann}, {McEnery}, {Mincer}, {Moskalenko}, {Nemethy},
  {Noyes}, {Porter}, {Pretz}, {Ryan}, {Parkinson}, {Shoup}, {Sinnis}, {Smith},
  {Strong}, {Sullivan}, {Vasileiou}, {Walker}, {Williams}, \&
  {Yodh}}]{2008ApJ...688.1078A}
{Abdo}, A.~A., {Allen}, B., {Aune}, T., {et~al.} 2008{\natexlab{a}}, \apj, 688,
  1078

\bibitem[{{Abdo} {et~al.}(2008{\natexlab{b}}){Abdo}, {Allen}, {Aune}, {Berley},
  {Blaufuss}, {Casanova}, {Chen}, {Dingus}, {Ellsworth}, {Fleysher},
  {Fleysher}, {Gonzalez}, {Goodman}, {Hoffman}, {H{\"u}ntemeyer}, {Kolterman},
  {Lansdell}, {Linnemann}, {McEnery}, {Mincer}, {Nemethy}, {Noyes}, {Pretz},
  {Ryan}, {Parkinson}, {Shoup}, {Sinnis}, {Smith}, {Sullivan}, {Vasileiou},
  {Walker}, {Williams}, \& {Yodh}}]{2008PhRvL.101v1101A}
---. 2008{\natexlab{b}}, Physical Review Letters, 101, 221101

\bibitem[{{Abdo} {et~al.}(2012){Abdo}, {Allen}, {Atkins}, {Aune}, {Benbow},
  {Berley}, {Blaufuss}, {Bonamente}, {Bussons}, {Chen}, {Christopher}, {Coyne},
  {DeYoung}, {Dingus}, {Dorfan}, {Ellsworth}, {Falcone}, {Fleysher},
  {Fleysher}, {Galbraith-Frew}, {Gonzalez}, {Goodman}, {Haines}, {Hays},
  {Hoffman}, {H{\"u}ntemeyer}, {Kelley}, {Kolterman}, {Lansdell}, {Linnemann},
  {McCullough}, {McEnery}, {Morgan}, {Mincer}, {Morales}, {Nemethy}, {Noyes},
  {Pretz}, {Ryan}, {Samuelson}, {Saz Parkinson}, {Shoup}, {Sinnis}, {Smith},
  {Sullivan}, {Vasileiou}, {Walker}, {Wascko}, {Williams}, {Westerhoff}, \&
  {Yodh}}]{2012ApJ...750...63A}
{Abdo}, A.~A., {Allen}, B.~T., {Atkins}, R., {et~al.} 2012, \apj, 750, 63

\bibitem[{{Acciari} {et~al.}(2011){Acciari}, {Aliu}, {Arlen}, {Aune},
  {Beilicke}, {Benbow}, {Boltuch}, {Bradbury}, {Buckley}, {Bugaev}, {Byrum},
  {Cannon}, {Cesarini}, {Ciupik}, {Cui}, {Dickherber}, {Duke}, {Falcone},
  {Finley}, {Finnegan}, {Fortson}, {Furniss}, {Galante}, {Gall}, {Gillanders},
  {Godambe}, {Grube}, {Guenette}, {Gyuk}, {Hanna}, {Holder}, {Hui}, {Humensky},
  {Imran}, {Kaaret}, {Karlsson}, {Kertzman}, {Kieda}, {Konopelko},
  {Krawczynski}, {Krennrich}, {Lang}, {Maier}, {McArthur}, {McCutcheon},
  {Moriarty}, {Ong}, {Otte}, {Ouellette}, {Pandel}, {Perkins}, {Pichel},
  {Pohl}, {Quinn}, {Ragan}, {Reyes}, {Reynolds}, {Roache}, {Rose}, {Rovero},
  {Schroedter}, {Sembroski}, {Senturk}, {Steele}, {Swordy}, {Theiling},
  {Thibadeau}, {Varlotta}, {Vassiliev}, {Vincent}, {Wagner}, {Wakely}, {Ward},
  {Weekes}, {Weinstein}, {Weisgarber}, {Williams}, {Wissel}, {Wood}, {Zitzer},
  {Garson}, {Lee}, {Sadun}, {Carini}, {Barnaby}, {Cook}, {Maune}, {Pease},
  {Smith}, {Walters}, {Berdyugin}, {Lindfors}, {Nilsson}, {Pasanen}, {Sainio},
  {Sillanpaa}, {Takalo}, {Villforth}, {Montaruli}, {Baker}, {Lahteenmaki},
  {Tornikoski}, {Hovatta}, {Nieppola}, {Aller}, \&
  {Aller}}]{2011ApJ...738...25A}
{Acciari}, V.~A., {Aliu}, E., {Arlen}, T., {et~al.} 2011, \apj, 738, 25

\bibitem[{{Aharonian} {et~al.}(2002){Aharonian}, {Akhperjanian}, {Beilicke},
  {Bernl{\"o}hr}, {B{\"o}rst}, {Bojahr}, {Bolz}, {Coarasa}, {Contreras},
  {Cortina}, {Costamante}, {Denninghoff}, {Fonseca}, {Girma}, {G{\"o}tting},
  {Heinzelmann}, {Hermann}, {Heusler}, {Hofmann}, {Horns}, {Jung}, {Kankanyan},
  {Kestel}, {Kettler}, {Kohnle}, {Konopelko}, {Kornmeyer}, {Kranich},
  {Krawczynski}, {Lampeitl}, {Lopez}, {Lorenz}, {Lucarelli}, {Mang}, {Meyer},
  {Mirzoyan}, {Milite}, {Moralejo}, {Ona}, {Panter}, {Plyasheshnikov},
  {P{\"u}hlhofer}, {Rauterberg}, {Reyes}, {Rhode}, {Ripken}, {Rowell},
  {Sahakian}, {Samorski}, {Schilling}, {Siems}, {Sobzynska}, {Stamm},
  {Tluczykont}, {V{\"o}lk}, {Wiedner}, {Wittek}, \&
  {Remillard}}]{2002A&A...393...89A}
{Aharonian}, F., {Akhperjanian}, A., {Beilicke}, M., {et~al.} 2002, \aap, 393,
  89

\bibitem[{{Aharonian} {et~al.}(2003){Aharonian}, {Akhperjanian}, {Beilicke},
  {Bernl{\"o}hr}, {B{\"o}rst}, {Bojahr}, {Bolz}, {Coarasa}, {Contreras},
  {Cortina}, {Denninghoff}, {Fonseca}, {Girma}, {Goebel}, {G{\"o}tting},
  {Heinzelmann}, {Hermann}, {Heusler}, {Hofmann}, {Horns}, {Jung}, {Kankanyan},
  {Kestel}, {Kettler}, {Kohnle}, {Konopelko}, {Kranich}, {Krawczynski},
  {Lampeitl}, {L{\'o}pez}, {Lorenz}, {Lucarelli}, {Mang}, {Meyer}, {Mirzoyan},
  {Moralejo}, {O{\~n}a-Wilhelmi}, {Paneque}, {Panter}, {Plyasheshnikov},
  {P{\"u}hlhofer}, {de los Reyes}, {Rhode}, {Ripken}, {Rowell}, {Sahakian},
  {Samorski}, {Schilling}, {Schweizer}, {Sevilla}, {Siems}, {Sobczy{\'n}ska},
  {Stamm}, {Tluczykont}, {Tonello}, {Vitale}, {V{\"o}lk}, {Wagner}, {Wiedner},
  \& {Wittek}}]{2003A&A...410..813A}
---. 2003, \aap, 410, 813

\bibitem[{{Aharonian} {et~al.}(2005){Aharonian}, {Akhperjanian}, {Aye},
  {Bazer-Bachi}, {Beilicke}, {Benbow}, {Berge}, {Berghaus}, {Bernl{\"o}hr},
  {Boisson}, {Bolz}, {Braun}, {Breitling}, {Brown}, {Bussons Gordo},
  {Chadwick}, {Chounet}, {Cornils}, {Costamante}, {Degrange},
  {Djannati-Ata{\"i}}, {O'C.~Drury}, {Dubus}, {Emmanoulopoulos}, {Espigat},
  {Feinstein}, {Fleury}, {Fontaine}, {Fuchs}, {Funk}, {Gallant}, {Giebels},
  {Gillessen}, {Glicenstein}, {Goret}, {Hadjichristidis}, {Hauser},
  {Heinzelmann}, {Henri}, {Hermann}, {Hinton}, {Hofmann}, {Holleran}, {Horns},
  {de Jager}, {Kh{\'e}lifi}, {Komin}, {Konopelko}, {Latham}, {Le Gallou},
  {Lemi{\`e}re}, {Lemoine}, {Leroy}, {Lohse}, {Marcowith}, {Masterson},
  {McComb}, {de Naurois}, {Nolan}, {Noutsos}, {Orford}, {Osborne}, {Ouchrif},
  {Panter}, {Pelletier}, {Pita}, {P{\"u}hlhofer}, {Punch}, {Raubenheimer},
  {Raue}, {Raux}, {Rayner}, {Redondo}, {Reimer}, {Reimer}, {Ripken}, {Rob},
  {Rolland}, {Rowell}, {Sahakian}, {Saug{\'e}}, {Schlenker}, {Schlickeiser},
  {Schuster}, {Schwanke}, {Siewert}, {Sol}, {Steenkamp}, {Stegmann},
  {Tavernet}, {Terrier}, {Th{\'e}oret}, {Tluczykont}, {Vasileiadis}, {Venter},
  {Vincent}, {V{\"o}lk}, \& {Wagner}}]{2005A&A...437...95A}
{Aharonian}, F., {Akhperjanian}, A.~G., {Aye}, K.-M., {et~al.} 2005, \aap, 437,
  95

\bibitem[{{Aielli} {et~al.}(2010){Aielli}, {Bacci}, {Bartoli}, {Bernardini},
  {Bi}, {Bleve}, {Branchini}, {Budano}, {Bussino}, {Calabrese Melcarne},
  {Camarri}, {Cao}, {Cappa}, {Cardarelli}, {Catalanotti}, {Cattaneo}, {Celio},
  {Chen}, {Chen}, {Cheng}, {Creti}, {Cui}, {Dai}, {D'Al{\'{\i}} Staiti},
  {Danzengluobu}, {Dattoli}, {De Mitri}, {D'Ettorre Piazzoli}, {De Vincenzi},
  {Di Girolamo}, {Ding}, {Di Sciascio}, {Feng}, {Feng}, {Feng}, {Galeazzi},
  {Galeotti}, {Gargana}, {Gou}, {Guo}, {He}, {Hu}, {Hu}, {Huang}, {Iacovacci},
  {Iuppa}, {James}, {Jia}, {Labaciren}, {Li}, {Li}, {Li}, {Liberti}, {Liguori},
  {Liu}, {Liu}, {Liu}, {Liu}, {Lu}, {Ma}, {Mancarella}, {Mari}, {Marsella},
  {Martello}, {Mastroianni}, {Meng}, {Montini}, {Ning}, {Pagliaro}, {Panareo},
  {Perrone}, {Pistilli}, {Qu}, {Rossi}, {Ruggieri}, {Saggese}, {Salvini},
  {Santonico}, {Shen}, {Sheng}, {Shi}, {Stanescu}, {Surdo}, {Tan}, {Vallania},
  {Vernetto}, {Vigorito}, {Wang}, {Wang}, {Wu}, {Wu}, {Xu}, {Xue}, {Yan},
  {Yang}, {Yang}, {Yuan}, {Zha}, {Zhang}, {Zhang}, {Zhang}, {Zhang}, {Zhang},
  {Zhang}, {Zhang}, {Zhaxisangzhu}, {Zhou}, {Zhu}, {Zhu}, {Zizzi}, \& {Argo-YBJ
  Collaboration}}]{2010ApJ...714L.208A}
{Aielli}, G., {Bacci}, C., {Bartoli}, B., {et~al.} 2010, \apjl, 714, L208

\bibitem[{{Aleksi{\'c}} {et~al.}(2010){Aleksi{\'c}}, {Anderhub}, {Antonelli},
  {Antoranz}, {Backes}, {Baixeras}, {Balestra}, {Barrio}, {Bastieri}, {Becerra
  Gonz{\'a}lez}, {Becker}, {Bednarek}, {Berdyugin}, {Berger}, {Bernardini},
  {Biland}, {Bock}, {Bonnoli}, {Bordas}, {Borla Tridon}, {Bosch-Ramon}, {Bose},
  {Braun}, {Bretz}, {Britzger}, {Camara}, {Carmona}, {Carosi}, {Colin},
  {Commichau}, {Contreras}, {Cortina}, {Costado}, {Covino}, {Dazzi}, {de
  Angelis}, {de Cea Del Pozo}, {de Los Reyes}, {de Lotto}, {de Maria}, {de
  Sabata}, {Delgado Mendez}, {Doert}, {Dom{\'{\i}}nguez}, {Dominis Prester},
  {Dorner}, {Doro}, {Elsaesser}, {Errando}, {Ferenc}, {Fonseca}, {Font},
  {Garc{\'{\i}}a L{\'o}pez}, {Garczarczyk}, {Gaug}, {Godinovic}, {Hadasch},
  {Herrero}, {Hildebrand}, {H{\"o}hne-M{\"o}nch}, {Hose}, {Hrupec}, {Hsu},
  {Jogler}, {Klepser}, {Kr{\"a}henb{\"u}hl}, {Kranich}, {La Barbera}, {Laille},
  {Leonardo}, {Lindfors}, {Lombardi}, {Longo}, {L{\'o}pez}, {Lorenz},
  {Majumdar}, {Maneva}, {Mankuzhiyil}, {Mannheim}, {Maraschi}, {Mariotti},
  {Mart{\'{\i}}nez}, {Mazin}, {Meucci}, {Miranda}, {Mirzoyan}, {Miyamoto},
  {Mold{\'o}n}, {Moles}, {Moralejo}, {Nieto}, {Nilsson}, {Ninkovic}, {Orito},
  {Oya}, {Paoletti}, {Paredes}, {Partini}, {Pasanen}, {Pascoli}, {Pauss},
  {Pegna}, {Perez-Torres}, {Persic}, {Peruzzo}, {Prada}, {Prandini},
  {Puchades}, {Puljak}, {Reichardt}, {Rhode}, {Rib{\'o}}, {Rico}, {Rissi},
  {R{\"u}gamer}, {Saggion}, {Saito}, {Salvati}, {S{\'a}nchez-Conde},
  {Satalecka}, {Scalzotto}, {Scapin}, {Schweizer}, {Shayduk}, {Shore},
  {Sierpowska-Bartosik}, {Sillanp{\"a}{\"a}}, {Sitarek}, {Sobczynska},
  {Spanier}, {Spiro}, {Stamerra}, {Steinke}, {Strah}, {Struebig}, {Suric},
  {Takalo}, {Tavecchio}, {Temnikov}, {Tescaro}, {Teshima}, {Torres}, {Vankov},
  {Wagner}, {Zabalza}, {Zandanel}, {Zanin}, \& {MAGIC
  Collaboration}}]{2010A&A...519A..32A}
{Aleksi{\'c}}, J., {Anderhub}, H., {Antonelli}, L.~A., {et~al.} 2010, \aap,
  519, A32

\bibitem[{{Aleksi{\'c}} {et~al.}(2012){Aleksi{\'c}}, {Alvarez}, {Antonelli},
  {Antoranz}, {Asensio}, {Backes}, {Barrio}, {Bastieri}, {Becerra
  Gonz{\'a}lez}, {Bednarek}, {Berdyugin}, {Berger}, {Bernardini}, {Biland},
  {Blanch}, {Bock}, {Boller}, {Bonnoli}, {Borla Tridon}, {Braun}, {Bretz},
  {Ca{\~n}ellas}, {Carmona}, {Carosi}, {Colin}, {Colombo}, {Contreras},
  {Cortina}, {Cossio}, {Covino}, {Dazzi}, {De Angelis}, {De Caneva}, {De Cea
  del Pozo}, {De Lotto}, {Delgado Mendez}, {Diago Ortega}, {Doert},
  {Dom{\'{\i}}nguez}, {Dominis Prester}, {Dorner}, {Doro}, {Elsaesser},
  {Ferenc}, {Fonseca}, {Font}, {Fruck}, {Garc{\'{\i}}a L{\'o}pez},
  {Garczarczyk}, {Garrido}, {Giavitto}, {Godinovi{\'c}}, {Hadasch},
  {H{\"a}fner}, {Herrero}, {Hildebrand}, {H{\"o}hne-M{\"o}nch}, {Hose},
  {Hrupec}, {Huber}, {Jogler}, {Kellermann}, {Klepser}, {Kr{\"a}henb{\"u}hl},
  {Krause}, {La Barbera}, {Lelas}, {Leonardo}, {Lindfors}, {Lombardi},
  {L{\'o}pez}, {L{\'o}pez}, {Lorenz}, {Makariev}, {Maneva}, {Mankuzhiyil},
  {Mannheim}, {Maraschi}, {Mariotti}, {Mart{\'{\i}}nez}, {Mazin}, {Meucci},
  {Miranda}, {Mirzoyan}, {Miyamoto}, {Mold{\'o}n}, {Moralejo}, {Munar-Adrover},
  {Nieto}, {Nilsson}, {Orito}, {Oya}, {Paneque}, {Paoletti}, {Pardo},
  {Paredes}, {Partini}, {Pasanen}, {Pauss}, {Perez-Torres}, {Persic},
  {Peruzzo}, {Pilia}, {Pochon}, {Prada}, {Prada Moroni}, {Prandini}, {Puljak},
  {Reichardt}, {Reinthal}, {Rhode}, {Rib{\'o}}, {Rico}, {R{\"u}gamer},
  {Saggion}, {Saito}, {Saito}, {Salvati}, {Satalecka}, {Scalzotto}, {Scapin},
  {Schultz}, {Schweizer}, {Shayduk}, {Shore}, {Sillanp{\"a}{\"a}}, {Sitarek},
  {Sobczynska}, {Spanier}, {Spiro}, {Stamerra}, {Steinke}, {Storz}, {Strah},
  {Suri{\'c}}, {Takalo}, {Takami}, {Tavecchio}, {Temnikov}, {Terzi{\'c}},
  {Tescaro}, {Teshima}, {Tibolla}, {Torres}, {Treves}, {Uellenbeck}, {Vankov},
  {Vogler}, {Wagner}, {Weitzel}, {Zabalza}, {Zandanel}, \&
  {Zanin}}]{2012A&A...542A.100A}
{Aleksi{\'c}}, J., {Alvarez}, E.~A., {Antonelli}, L.~A., {et~al.} 2012, \aap,
  542, A100

\bibitem[{{Atkins} {et~al.}(2004){Atkins}, {Benbow}, {Berley}, {Blaufuss},
  {Bussons}, {Coyne}, {De Young}, {Dingus}, {Dorfan}, {Ellsworth}, {Fleysher},
  {Fleysher}, {Gisler}, {Gonzalez}, {Goodman}, {Haines}, {Hays}, {Hoffman},
  {Kelley}, {Lansdell}, {Linnemann}, {McEnery}, {Miller}, {Mincer}, {Morales},
  {Nemethy}, {Noyes}, {Ryan}, {Samuelson}, {Shoup}, {Sinnis}, {Smith},
  {Sullivan}, {Williams}, {Westerhoff}, {Wilson}, {Xu}, \&
  {Yodh}}]{2004ApJ...608..680A}
{Atkins}, R., {Benbow}, W., {Berley}, D., {et~al.} 2004, \apj, 608, 680

\bibitem[{{Bartoli} {et~al.}(2011){Bartoli}, {Bernardini}, {Bi}, {Bleve},
  {Bolognino}, {Branchini}, {Budano}, {Calabrese Melcarne}, {Camarri}, {Cao},
  {Cappa}, {Cardarelli}, {Catalanotti}, {Cattaneo}, {Celio}, {Chen}, {Chen},
  {Chen}, {Creti}, {Cui}, {Dai}, {D'Al{\'{\i}} Staiti}, {Danzengluobu},
  {Dattoli}, {De Mitri}, {D'Ettorre Piazzoli}, {Di Girolamo}, {Ding}, {Di
  Sciascio}, {Feng}, {Feng}, {Feng}, {Galeazzi}, {Galeotti}, {Giroletti},
  {Gou}, {Guo}, {He}, {Hu}, {Hu}, {Huang}, {Iacovacci}, {Iuppa}, {James},
  {Jia}, {Labaciren}, {Li}, {Li}, {Li}, {Liguori}, {Liu}, {Liu}, {Liu}, {Liu},
  {Lu}, {Ma}, {Mancarella}, {Mari}, {Marsella}, {Martello}, {Mastroianni},
  {Montini}, {Ning}, {Pagliaro}, {Panareo}, {Panico}, {Perrone}, {Pistilli},
  {Qu}, {Rossi}, {Ruggieri}, {Salvini}, {Santonico}, {Shen}, {Sheng}, {Shi},
  {Stanescu}, {Surdo}, {Tan}, {Vallania}, {Vernetto}, {Vigorito}, {Wang},
  {Wang}, {Wu}, {Wu}, {Xu}, {Xue}, {Yan}, {Yang}, {Yang}, {Yao}, {Yuan}, {Zha},
  {Zhang}, {Zhang}, {Zhang}, {Zhang}, {Zhang}, {Zhang}, {Zhang}, {Zhaxiciren},
  {Zhaxisangzhu}, {Zhou}, {Zhu}, {Zhu}, {Zizzi}, \& {ARGO-YBJ
  Collaboration}}]{2011ApJ...734..110B}
{Bartoli}, B., {Bernardini}, P., {Bi}, X.~J., {et~al.} 2011, \apj, 734, 110

\bibitem[{{B{\l}a{\.z}ejowski} {et~al.}(2005){B{\l}a{\.z}ejowski}, {Blaylock},
  {Bond}, {Bradbury}, {Buckley}, {Carter-Lewis}, {Celik}, {Cogan}, {Cui},
  {Daniel}, {Duke}, {Falcone}, {Fegan}, {Fegan}, {Finley}, {Fortson},
  {Gammell}, {Gibbs}, {Gillanders}, {Grube}, {Gutierrez}, {Hall}, {Hanna},
  {Holder}, {Horan}, {Humensky}, {Kenny}, {Kertzman}, {Kieda}, {Kildea},
  {Knapp}, {Kosack}, {Krawczynski}, {Krennrich}, {Lang}, {LeBohec}, {Linton},
  {Lloyd-Evans}, {Maier}, {Mendoza}, {Milovanovic}, {Moriarty}, {Nagai}, {Ong},
  {Power-Mooney}, {Quinn}, {Quinn}, {Ragan}, {Reynolds}, {Rebillot}, {Rose},
  {Schroedter}, {Sembroski}, {Swordy}, {Syson}, {Valcarel}, {Vassiliev},
  {Wakely}, {Walker}, {Weekes}, {White}, {Zweerink}, {VERITAS Collaboration},
  {Mochejska}, {Smith}, {Aller}, {Aller}, {Ter{\"a}sranta}, {Boltwood},
  {Sadun}, {Stanek}, {Adams}, {Foster}, {Hartman}, {Lai}, {B{\"o}ttcher},
  {Reimer}, \& {Jung}}]{2005ApJ...630..130B}
{B{\l}a{\.z}ejowski}, M., {Blaylock}, G., {Bond}, I.~H., {et~al.} 2005, \apj,
  630, 130

\bibitem[{{Cui}(2004)}]{2004ApJ...605..662C}
{Cui}, W. 2004, \apj, 605, 662

\bibitem[{{de Vaucouleurs} {et~al.}(1991){de Vaucouleurs}, {de Vaucouleurs},
  {Corwin}, {Buta}, {Paturel}, \& {Fouqu{\'e}}}]{1991rc3..book.....D}
{de Vaucouleurs}, G., {de Vaucouleurs}, A., {Corwin}, Jr., H.~G., {et~al.}
  1991, {Third Reference Catalogue of Bright Galaxies. Volume I: Explanations
  and references. Volume II: Data for galaxies between 0$^{h}$ and 12$^{h}$.
  Volume III: Data for galaxies between 12$^{h}$ and 24$^{h}$.}

\bibitem[{{Fossati} {et~al.}(2008){Fossati}, {Buckley}, {Bond}, {Bradbury},
  {Carter-Lewis}, {Chow}, {Cui}, {Falcone}, {Finley}, {Gaidos}, {Grube},
  {Holder}, {Horan}, {Horns}, {Jordan}, {Kieda}, {Kildea}, {Krawczynski},
  {Krennrich}, {Lang}, {LeBohec}, {Lee}, {Moriarty}, {Ong}, {Petry}, {Quinn},
  {Sembroski}, {Wakely}, \& {Weekes}}]{2008ApJ...677..906F}
{Fossati}, G., {Buckley}, J.~H., {Bond}, I.~H., {et~al.} 2008, \apj, 677, 906

\bibitem[{{Helene}(1983)}]{1983NIMPR.212..319H}
{Helene}, O. 1983, Nuclear Instruments and Methods in Physics Research, 212,
  319

\bibitem[{{Konopelko} {et~al.}(2008){Konopelko}, {Cui}, {Duke}, \&
  {Finley}}]{2008ApJ...679L..13K}
{Konopelko}, A., {Cui}, W., {Duke}, C., \& {Finley}, J.~P. 2008, \apjl, 679,
  L13

\bibitem[{{Krawczynski} {et~al.}(2004){Krawczynski}, {Hughes}, {Horan},
  {Aharonian}, {Aller}, {Aller}, {Boltwood}, {Buckley}, {Coppi}, {Fossati},
  {G{\"o}tting}, {Holder}, {Horns}, {Kurtanidze}, {Marscher}, {Nikolashvili},
  {Remillard}, {Sadun}, \& {Schr{\"o}der}}]{2004ApJ...601..151K}
{Krawczynski}, H., {Hughes}, S.~B., {Horan}, D., {et~al.} 2004, \apj, 601, 151

\bibitem[{{Krennrich} {et~al.}(2001){Krennrich}, {Badran}, {Bond}, {Bradbury},
  {Buckley}, {Carter-Lewis}, {Catanese}, {Cui}, {Dunlea}, {Das}, {de la Calle
  Perez}, {Fegan}, {Fegan}, {Finley}, {Gaidos}, {Gibbs}, {Gillanders}, {Hall},
  {Hillas}, {Holder}, {Horan}, {Jordan}, {Kertzman}, {Kieda}, {Kildea},
  {Knapp}, {Kosack}, {Lang}, {LeBohec}, {McKernan}, {Moriarty}, {M{\"u}ller},
  {Ong}, {Pallassini}, {Petry}, {Quinn}, {Reay}, {Reynolds}, {Rose},
  {Sembroski}, {Sidwell}, {Stanton}, {Swordy}, {Vassiliev}, {Wakely}, \&
  {Weekes}}]{2001ApJ...560L..45K}
{Krennrich}, F., {Badran}, H.~M., {Bond}, I.~H., {et~al.} 2001, \apjl, 560, L45

\bibitem[{{Kusunose} \& {Takahara}(2006)}]{2006ApJ...651..113K}
{Kusunose}, M., \& {Takahara}, F. 2006, \apj, 651, 113

\bibitem[{{Padovani} \& {Giommi}(1995)}]{1995ApJ...444..567P}
{Padovani}, P., \& {Giommi}, P. 1995, \apj, 444, 567

\bibitem[{{Patricelli} {et~al.}(\noop{3002}2013a){Patricelli}, {Gonz\'alez},
  {Marinelli}, \& {the Milagro collaboration}}]{RICAPProc}
{Patricelli}, B., {Gonz\'alez}, M.M., F.~N., {Marinelli}, A., \& {the Milagro
  collaboration}. \noop{3002}2013a, NIMA

\bibitem[{{Patricelli} {et~al.}(\noop{3002}2013b, to be published){Patricelli},
  {Gonz\'alez}, {Marinelli}, \& {the Milagro collaboration}}]{IcrcProc}
{Patricelli}, B., {Gonz\'alez}, M.M., F.~N., {Marinelli}, A., \& {the Milagro
  collaboration}. \noop{3002}2013b, to be published, in the Proceedings of the
  ICRC 2013 Meeting in Rio de Janeiro, Brazil, July 2-9, 2013

\bibitem[{{Rebillot} {et~al.}(2006){Rebillot}, {Badran}, {Blaylock},
  {Bradbury}, {Buckley}, {Carter-Lewis}, {Celik}, {Chow}, {Cogan}, {Cui},
  {Daniel}, {Duke}, {Falcone}, {Fegan}, {Finley}, {Fortson}, {Gillanders},
  {Grube}, {Gutierrez}, {Gyuk}, {Hanna}, {Holder}, {Horan}, {Hughes}, {Kenny},
  {Kertzman}, {Kieda}, {Kildea}, {Kosack}, {Krawczynski}, {Krennrich}, {Lang},
  {Le Bohec}, {Linton}, {Maier}, {Moriarty}, {Perkins}, {Pohl}, {Quinn},
  {Ragan}, {Reynolds}, {Rose}, {Schroedter}, {Sembroski}, {Steele}, {Swordy},
  {Valcarcel}, {Vassiliev}, {Wakely}, {Weekes}, {Zweerink}, {VERITAS
  Collaboration}, {Aller}, {Aller}, {Boltwood}, {Jung}, {Kranich}, {Nilsson},
  {Pasanen}, {Sadun}, \& {Sillanpaa}}]{2006ApJ...641..740R}
{Rebillot}, P.~F., {Badran}, H.~M., {Blaylock}, G., {et~al.} 2006, \apj, 641,
  740

\bibitem[{{Reimer} {et~al.}(2005){Reimer}, {B{\"o}ttcher}, \&
  {Postnikov}}]{2005ApJ...630..186R}
{Reimer}, A., {B{\"o}ttcher}, M., \& {Postnikov}, S. 2005, \apj, 630, 186

\bibitem[{{Resconi} {et~al.}(2009){Resconi}, {Franco}, {Gross}, {Costamante},
  \& {Flaccomio}}]{2009A&A...502..499R}
{Resconi}, E., {Franco}, D., {Gross}, A., {Costamante}, L., \& {Flaccomio}, E.
  2009, \aap, 502, 499

\bibitem[{{Sahu} {et~al.}(2013){Sahu}, {Oliveros}, \&
  {Sanabria}}]{2013PhRvD..87j3015S}
{Sahu}, S., {Oliveros}, A.~F.~O., \& {Sanabria}, J.~C. 2013, \prd, 87, 103015

\bibitem[{{Tluczykont} {et~al.}(2010){Tluczykont}, {Bernardini}, {Satalecka},
  {Clavero}, {Shayduk}, \& {Kalekin}}]{2010A&A...524A..48T}
{Tluczykont}, M., {Bernardini}, E., {Satalecka}, K., {et~al.} 2010, \aap, 524,
  A48

\bibitem[{{Tluczykont} {et~al.}(2007){Tluczykont}, {Shayduk}, {Kalekin}, \&
  {Bernardini}}]{2007JPhCS..60..318T}
{Tluczykont}, M., {Shayduk}, M., {Kalekin}, O., \& {Bernardini}, E. 2007,
  Journal of Physics Conference Series, 60, 318

\bibitem[{{Wagner}(2008)}]{2008MNRAS.385..119W}
{Wagner}, R.~M. 2008, \mnras, 385, 119

\end{thebibliography}
